# The Quantum Twisting Microscope

A. Inbar[1†], J. Birkbeck[1†*], J. Xiao[1†], T. Taniguchi[2], K. Watanabe[2],

B. Yan[1], Y. Oreg[1], Ady Stern[1], E. Berg[1], and S. Ilani[1]

[1] *Department of Condensed Matter Physics, Weizmann Institute of Science, Rehovot 76100, Israel.*
[2] *National Institute for Materials Science, 1-1 Namiki, Tsukuba, 305-0044 Japan.*
† These authors contributed equally to the work.
* Correspondence to: john.birkbeck@weizmann.ac.il

The invention of scanning probe microscopy has revolutionized the way electronic phenomena are visualized. While present-day probes can access a variety of electronic properties at a single location in space, a scanning microscope that can directly probe the quantum mechanical existence of an electron at multiple locations would provide direct access to key quantum properties of electronic systems, so far unreachable. Here, we demonstrate a conceptually new type of scanning probe microscope – the Quantum Twisting Microscope (QTM) – capable of performing local interference experiments at its tip. The QTM is based on a unique van-der-Waals tip, allowing the creation of pristine 2D junctions, which provide a multitude of coherently-interfering paths for an electron to tunnel into a sample. With the addition of a continuously scanned twist angle between the tip and sample, this microscope probes electrons in momentum space similar to the way a scanning tunneling microscope probes electrons in real space. Through a series of experiments, we demonstrate room temperature quantum coherence at the tip, study the twist angle evolution of twisted bilayer graphene, directly image the energy bands of monolayer and twisted bilayer graphene, and finally, apply large local pressures while visualizing the evolution of the flat energy band of the latter. The QTM opens the way for novel classes of experiments on quantum materials.

An electron in a solid is a quantum mechanical object best described by an extended wave function, reflecting its existence in a superposition of spatial locations. The invention of the scanning tunneling microscope (STM)[1] dramatically changed how we can visualize electrons in real space. It was followed by a large array of other scanning probe techniques that now allow local measurements of various electronic properties[2]. To date, existing scanning microscopes can only probe electronic properties at one location at a time. Therefore they are unable to probe the interference of several tunneling trajectories, which would map the evolution of the quantum mechanical phase in space. To do so requires a scanning interferometer that probes an electron at several locations simultaneously and quantum coherently.

In this work, we demonstrate a conceptually new type of scanning probe microscope – the Quantum Twisting Microscope (QTM) – capable of performing local interference experiments at its tip. Our technique is based on a unique van-der-Waals (vdW) tip, which is brought into contact with a vdW sample, forming a 2D interface that can be twisted and spatially scanned with high angular and positional precision. The QTM enables two orthogonal and complementary classes of experiments: 1) "In-situ twistronics" – measuring a continuously twistable interface between a pair of vdW materials (e.g. twisted bilayer graphene[3–5]). Here, we place the active vdW layers in the tip and sample in direct contact, such that their wavefunctions strongly couple, and probe the transport properties of the hybridized interface. 2) Momentum-resolved tunneling – Here, we insert a tunnel barrier (e.g. a few layers of a transition metal dichalcogenide) between the active vdW layers in the tip and sample. This barrier decouples their wavefunctions, allowing the tip to act as a non-invasive momentum-resolved probe for the sample's energy bands. Since an electron can coherently tunnel into a sample at many locations along the 2D interface, this junction acts as an interferometer on a tip. Specifically, this implies that in the absence of electron-phonon, electron-electron and impurity scattering, an electron would tunnel only between states of equal momenta[6–16]. In this modality, the microscope's twisting degree of freedom is used for scanning an arc in the momentum space of the sample, and imaging along it the sample's energy bands.

**The QTM's Operating Principle**

To date, existing approaches for in-situ twisting experiments are based on electrical devices fabricated with a rotatable part and an external mechanical device, such as an atomic force microscope (AFM) tip, which pushes this part [17–21]. Our QTM, in contrast, elevates the AFM tip to become an integral part of the twisted device, which is now split into two parts: the first is a standard vdW device formed on a flat substrate (Fig. 1a, Supp. Info. S2). The second is a vdW device formed on a specially-designed pyramid at the edge of an AFM cantilever (Fig. 1b, Supp. Info. S1). Both sides have independent electrical contacts. We use a commercial AFM to bring the two parts into contact and to maintain a constant force across the interface throughout the entire experiment (Fig. 1c). On the AFM stage, we mount a piezoelectric rotator with X and Y nanopositioners on top (Supp. Info. S3). This setup allows rotating the bottom sample with an angular resolution of 0.001° and positioning the point of interest in the sample at the center of rotation. In addition, the standard scanning capability of the AFM (in X and Y directions) enables lateral scanning of the tip across the sample.

A crucial ingredient of the QTM is its tip design, facilitating the formation of a flat vdW plateau at its apex. To achieve this, we start with focused-ion-beam deposition of a platinum pyramid ($\sim 1.2 - 1.6 \mu m$ tall) on a tipless AFM cantilever (Fig. 1e). This is followed by sequentially transferring graphite, hBN and the active vdW layer (e.g., monolayer graphene) on the pyramid using a polymer membrane[22]. The graphite screens the substrate's disorder potential, and the hBN acts as a spacer. Fig. 1f shows an AFM image of the resulting tip: visibly, the vdW stack forms a "tent" over the pyramid with three or more folds climbing up to the pyramid's apex. At the apex, a flat plateau spontaneously forms in the vdW stack, whose corners are determined by the folds (Fig. 1g). The typical bending angles of the vdW tent ($\sim 10 - 30°$) are smaller than the pyramid's angle ($\sim 45°$). Thus, apart from a small touching point at the apex, the tent is mostly suspended (Supp. Info. S1). Due to the flexural rigidity of the graphite/hBN layers, the formed vdW plateau is wider than the pyramid apex, resting on it as a pivoting point. Therefore, when this tip is brought into contact with the sample, the plateau self-aligns its tilt to become parallel to the sample. By varying the pyramid geometry and graphite/hBN flake thicknesses (typically tens of nanometers), we achieve plateaus of varying linear dimensions between 50nm and 1μm. These dimensions are small enough to make our QTM tip a local scanning probe, yet large enough (~1000 atoms across or more) for wavefunctions on the plateau to have a well-defined momentum ($\Delta P \sim 1/1000$ of the Brillouin zone size). Contrary to etched boundaries in lithographic devices, the active vdW layer on the plateau is continuously extended to the rest of the pyramid, avoiding dangling bonds or buckles. This setting makes the wavefunctions smoothly connect to the rest of the vdW layer on the pyramid. Fig. 1d overlays the measurement circuit on a schematic cross-section of the junction: a bias voltage, $V_b$, is applied between the two active layers (vdW crystals 1 and 2), and the corresponding current, $I$, is measured. Buried bottom and top graphite layers can serve additionally as bottom and top gates for the junction.

**In-Situ Twistronics**

We start with a "twistronic" measurement of a twistable interface between two graphene monolayers (MLG) in direct contact. This simple interface has remained elusive for existing in-situ twistronics techniques[17–19] since MLG easily crumbles upon twisting. Fig. 1h shows the MLG-MLG interface conductance, $dI/dV$, measured vs. twist angle, $\theta$, at T = 300K. Throughout the entire measurement, the sample and tip are kept in continuous contact. The layers do not show any sign of locking at $\theta = 0º$, and the conductance traces are highly reproducible, highlighting the unique mechanical robustness of the QTM junction. The conductance is mirror symmetric around $\theta = 30°$, ($\theta \rightarrow 60° - \theta$), where its value is minimal. It rises continuously toward $\theta = 0°$ but plateaus at small angles ($|\theta| \lesssim 4°$) where it becomes limited by the resistance away from the tip ("contact resistance"). Similar to the measurements in graphite-graphite[17] and graphite-MLG[18] interfaces, we see extremely sharp conductance peaks at $\theta = 21.8º$ and 38.2º. At these angles, the two layers form commensurate stackings in real-space[17,18,23,24] with a $\sqrt{7} \times \sqrt{7}$ supercell (Fig. 1i).

What is the origin of the large conductance enhancement at commensurate angles? One possibility is a better real-space registry of the atoms in the two layers. However, by definition, whenever unit cells are commensurate in real-space, their corresponding Brillouin zones (BZ) are also commensurate in k-space, implying that at commensurate angles, momentum states are also matched. Specifically, at $\theta = 0º$ the Dirac cones of the two layers overlap at the corners of the 1$^{st}$ BZ, and at $\theta =21.8º$ they overlap at the corners of the 3rd BZ (Fig. 1i). The relevance of real vs. momentum space matching is directly connected to the quantum coherence of the 2D tunnel junction: in an incoherent junction, the tunneling events of electrons at different locations are independent, and sum up classically to yield the total current (Fig. 1j, top). In such a case only the real-space matching is relevant. Conversely, in a coherent junction, tunneling events at different locations interfere, yielding a tunneling current that is sensitive also to the local phases of the wavefunctions (Fig. 1j, bottom). In particular, tunneling is possible only between wavefunctions with matching energy and momentum.

**Local Momentum Resolved Tunneling**

To observe the momentum conserving nature of the QTM we add a tunneling barrier between the two MLG layers (trilayer WSe$_2$, Fig. 2a, top inset). This barrier suppresses the hybridization between tip and sample, enabling the tip to act as a probe for the unperturbed energy bands of the sample. The barrier also significantly increases the tunnel junction resistance, assuring that an applied bias falls predominantly across this junction and that the measurement is not affected by contact resistance even near $\theta = 0°$. Fig. 2a shows the measured tunneling current, $I$, vs. the interlayer bias, $V_b$, and $\theta$ at T = 300K. Around $\theta = 0°$, $I$ increases slowly with $V_b$ at low bias, and then sharply increases along a curved-X feature in the $\theta - V_b$ plane. Interestingly, this increase is followed by a sharp drop at slightly higher $V_b$. At much higher biases ($\sim 0.8V$), $I$ rises again, this time exponentially with $V_b$, and rather homogenously for all $\theta$. Fig. 2b shows the measured conductance, $dI/dV$. Here, the sharp drop of $I$ manifests as a strong negative differential resistance (NDR), as was seen previously in lithographic devices[9–12,14,16]. Finer details are revealed when we plot the second derivative, $d^2I/dV^2$ (Fig. 2c): in addition to the strong curved-X (dashed white lines), we observe a straight-X feature (dashed black lines), along which $d^2I/dV^2$ shows peaks (/dips) on the positive (/negative) bias side.

To understand the observed curved-X feature, consider three points in the $\theta - V_b$ plane (Fig. 2d): at point 1 ($\theta = 0°$, small $V_b$) the atoms of the two layers are registered in real-space and their energy surfaces are matched in momentum-space (Fig. 2g, panel 1). Increasing $V_b$ while keeping $\theta = 0°$ (point 2) maintains the real-space registry but offsets the relative energies of the Dirac cones. Consequently, for almost all energy slices within the bias window, the equal energy states have mismatched momenta (we show one such slice in panel 2 of Fig. 2g). Thus, the observed drop of $I$ is concomitant with the loss of energy-momentum-matched eigenstates in the two layers. By twisting the layers to a finite $\theta$, while maintaining $V_b$ (point 3), the layers lose the registry in real space but regain overlap of states in momentum space, enabling momentum-conserving tunneling processes (Fig. 2g panel 3). Indeed, at this point the measured $I$ becomes large again, demonstrating its momentum conserving nature. We estimate the level of momentum conservation from a plot of $I$ vs. $\theta$ at small $V_b = 40 mV$ (Fig. 2a, right inset). The extremely narrow peak ($\Delta\theta \approx 0.2°$) implies an excellent momentum resolution, ~0.004 of the BZ size, comparable to state-of-the-art ARPES detectors[25,26]. In the experiment in Fig. 2, the back gate voltage is zero and the measurement shows a good symmetry between positive and negative bias directions, suggesting an overall charge-neutral system. Additional experiments with finite back gate voltages (Supp. Info. S6) further show a rich evolution of the features with the total charge in the system.

We compare our measurements to theory of momentum-resolved tunneling between twisted layers[23] (Fig 2d-f). Since the quantum and geometrical capacitances of the junction are generally comparable, the applied bias divides into shifts of the chemical potential of the top and bottom layers, $\mu_T$ and $\mu_B$, and an electrostatic potential shift between them, $\phi$, namely, $V_b = \phi + \mu_B - \mu_T$ (Fig. 2h). To calculate the tunneling current at any point in the $\theta - V_b$ plane, we first determine $\mu_T$, $\mu_B$ and $\phi$ by solving the above equation self-consistently with the equations-of-state of the individual layers, $\mu_T(n_T)$ and $\mu_B(n_B)$, (Supp. Info. S4). We then sum the tunneling rates for all energy-momentum conserving tunneling processes within the bias window. We further add a Nordheim-Fowler[27] contribution due to the breakdown of the WSe$_2$ barrier. All expressions include the effects of finite temperature and lifetime. Overall, the agreement with the experiments is excellent, both in terms of the locations of the various features, and in terms of their relative magnitude. The theory further allows us to identify the experimentally observed features: the straight-X corresponds to the onset of momentum resolved tunneling, happening when the Fermi surface of one layer touches the empty bands of the other layer ('k-resolved onset' condition, Fig. 2f). The curved-X feature corresponds to nesting of the energy bands ('nesting' condition, Fig. 2f). Here, a macroscopically large number of energy-momentum conserving states become available to tunnel, explaining the large increase of $I$ along this feature (Supp. Video 4). With a further increase of $V_b$ or a further decrease of $\theta$ most of the states cease to be energy-momentum conserving, explaining the consequent large drop in $I$.

The simplicity of graphene's dispersion allows us to obtain analytic expressions for the k-resolved onset and nesting conditions. Specifically, for a non-interacting charge-neutral system, the onset condition becomes (Supp. Info S5) $V_b = \hbar v_F K_D \theta$, where $K_D$ is the Dirac point momentum. Namely, $V_b$ and $\theta$ substitute the energy and momentum in the standard Dirac equation. Overlaying this expression on the measurement (dashed black lines, Fig. 2c) yields an excellent fit with $v_F = 1.05 \pm 0.02 \cdot 10^6 m/s$, consistent with the measured Fermi velocity of graphene[28]. Additionally, the nesting line provides the energy shift between the bands, $\phi$, at any value of $V_b$, from which the electronic compressibility of the system can be straightforwardly determined (Supp. Info. S5). This measurement thus provides simultaneous information about the excitation spectrum of the system and its thermodynamic properties.

**Momentum-resolved Imaging of Twisted Bilayer Graphene Energy Bands**

Having probed the energy bands of MLG, we now turn to a system with more intricate energy bands – twisted bilayer graphene (TBG). The experiment comprises of a MLG probe, a bilayer Wse₂ barrier, and TBG with a 2.7° twist (Fig. 3a). In momentum space, the TBG mini-BZ hosts at its corners the Dirac cones of the underlying top and bottom graphene sheets (red and blue circles at $K_{top}$ and $K_{bot}$, Fig. 3b). In the experiment, the MLG is rotated with respect to the TBG, and correspondingly its Dirac cone (purple circle, Fig. 3b) scans the energy bands of the TBG along a constant radius arc in momentum space (dashed purple arc, Fig. 3b), cutting precisely through $K_{top}$ and $K_{bottom}$ and passing very close to the $\Gamma_M$ points of adjacent mini-BZ's. Along this k-space linecut, the TBG is theoretically predicted to exhibit the "flat bands" around zero energy and remote bands at higher energies[3,29] (blue and black in Fig. 3c). Fig. 3d shows the $dI^2/dV^2$ measured vs. $V_b$ and $\theta$. The second derivative diminishes the smoothly evolving background, allowing to clearly resolve the key features. The measurement shows a wealth of features, the most prominent of which are traced in Fig. 3f. Interestingly, it exhibits a superposition of the 'system' and 'probe' energy bands, including the TBG's flat bands (blue), remote bands (black), as well as two copies of the MLG Dirac bands (purple).

A theoretical calculation of the MLG–TBG momentum-resolved tunneling (Fig. 3e) shows excellent agreement with the experiment, down to small details. Specifically, the theory reproduces the features due to the flat and remote TBG bands, as well as the two displaced copies of the MLG Dirac bands. Furthermore, from the theory we can conclude that the band imaging is performed effectively by Dirac points: for example, when the MLG Dirac point matches in energy and momentum a state on the TBG flat bands (left inset, Fig. 3e), the $d^2I/dV^2$ exhibits a peak. This peak of $d^2I/dV^2$ corresponds to a minimum in $I$, reflecting the minimal tunneling density of states at the Dirac point. Similarly, when one of the two TBG's Dirac points match the MLG bands (right inset, Fig. 3e) a strong feature appears in $dI^2/dV^2$ (this time with an opposite sign, see toy model in Supp. Info S8). Consequently, the two TBG Dirac points trace out two copies of the MLG bands, displaced by $\Delta\theta = 2.7°$. In addition to these 'Dirac matching' conditions that are insensitive to band filling, there are also 'onset' features, which appear whenever the Fermi level of one side (tip/sample) crosses the bands of the other side (for details see Supp. Info. S8).

We obtain the flat-band energy dispersion directly from the measurements by using the simultaneously measured MLG Dirac bands to calibrate the energy shift at each $V_b$ (Supp. Info. S8). Fig. 3h compares the extracted $\epsilon(k)$ of the flat bands (red and grey dots) with the prediction of the BM model (blue). The overall agreement is rather good, although the experiment shows an electron-hole asymmetry (~12% along the $K_{bot} - M - K_{top}$ branch and ~20% along the $K_{top} - \Gamma_M$ branch), contrasting the nearly-perfect e-h symmetry of in the BM model. Our measurements also reveal a quantity that is inaccessible to other probes – the layer polarization of individual wavefunctions at different momenta: visibly, when the MLG lattice is rotationally aligned with the top TBG layer ($\theta = +1.35°$) the measured $dI^2/dV^2$ amplitudes are strong. In contrast, when the MLG is rotationally aligned with the bottom TBG layer ($\theta = -1.35°$) they are substantially weaker. This highlights the fact that the experiment probes the weight of the wavefunction on the top layer. In Fig 3i we plot the magnitude of $I$ along the traced flat band features (blue, Fig. 3f) between $K_{top}$ and $K_{bot}$ (dots), and compare it with the layer polarization as a function of $k$ from the BM model. We can see that this quantity is indeed a good proxy for the layer polarization.

**Imaging Twisted Bilayer Graphene Energy Bands Under Pressure**

We end by showing another unique capability of the QTM – its ability to apply a large local pressure while simultaneously imaging the way it affects the energy bands. Pressure provides a key tuning parameter for vdW materials[30–34], as it directly controls the interlayer tunneling. In TBG, this was predicted[30,33,34] and shown[32] to tune the flat bands in and out of the magic-angle condition. However, the challenges in cryogenic pressure cell transport experiments have limited such experiments to very few measurements. Here, instead, we use the ability of the AFM to apply $\mu N$-scale forces across the small QTM junction to achieve GPa-scale pressures at the interface (Fig 4a, inset).

Fig. 4a-c plots the measured $dI^2/dV^2$ vs. $\theta$ and $V_b$, for the junction of Fig. 3, but now under pressures of $P = 0.01, 0.4,$ and $0.68\ GPa$. Each frame is taken with a constant pressure during twisting, and the results are reversible upon repeatable increase and decrease of pressure between frames. Visibly, with increasing $P$ the flat bands gradually shrink toward zero $V_b$, whereas the remote bands get further away from zero $V_b$ (in more details, Supp. Video 1). To show this more quantitatively, in Fig 4d we trace the flat and remote bands over a larger sequence of pressures. We can clearly see the opposite motion of the flat and remote bands with pressure (arrows). This contrasting response reflects a band anti-crossing that increases with $P$, as expected from increased interlayer tunneling. Converting $V_b$ to the energy shift using the simultaneously measured MLG bands, we plot the energetic width of the flat bands vs. $P$ (Fig. 4e). Notably, the width shrinks linearly with P, reaching a 17% reduction at $P = 0.68\ GPa$. This compares reasonably to the 6-14% reduction predicted theoretically[30,33,34]. A naïve linear extrapolation of our measurements would suggest that the bands of 2.7° TBG could become fully flat at $P \approx 4 GPa$, well within the pressures achievable by AFM without damaging graphene[35,36]. This could lead to fully flat bands with a moiré periodicity that is much shorter than in magic-angle TBG, and correspondingly to proportionally larger Coulomb interactions, potentially taking this system into uncharted regimes of strong interactions.

The QTM demonstrated here opens the way for two independent research directions: In the first, it provides a new approach for creating highly controllable novel interfaces between a large variety of quantum materials. Specifically, it enables continuous control with 0.001° resolution of one of the critical parameters of these interfaces – their twist angle. In the current paper, we explored systems based on graphene and $WSe_2$, but the technique should be applicable quite generally to a plethora of layered conductors[4,37–41], semiconductors[41,42], and superconductors[43–45]. In the second, it is a novel scanning microscope that has direct access to the energy-momentum dispersion of electronic systems. As such, it may probe the dispersion of any excitation, charged or neutral, as long as it can be excited by a tunneling electron. The measurements can be performed at large magnetic fields, with variable carrier density and electrical displacement fields controlled by local gates, and with a continuously tunable pressure. In this manuscript we did not discuss the lateral scanning degrees of freedom of the QTM, which result naturally from its AFM platform. These degrees of freedom will further enable preforming spatially-scanned momentum resolved measurements within electronic devices with a high spatial resolution (~100nm). Lateral scanning will also provide control over the lateral displacements between vdW materials, most likely down to atomic dimensions, capturing another key tunning parameter of the energy dispersions at these interfaces. Given the relative simplicity of the technique and its powerful capabilities, we expect the QTM to become a valuable new tool in the arsenal of experimental condensed matter physics.

44. Farrar, L. S. *et al.* Superconducting Quantum Interference in Twisted van der Waals Heterostructures. *Nano Letters* **21**, 6725–6731 (2021).
45. Zhao, S. Y. F. *et al.* Emergent Interfacial Superconductivity between Twisted Cuprate Superconductors. *arXiv:2108.13455* (2021).

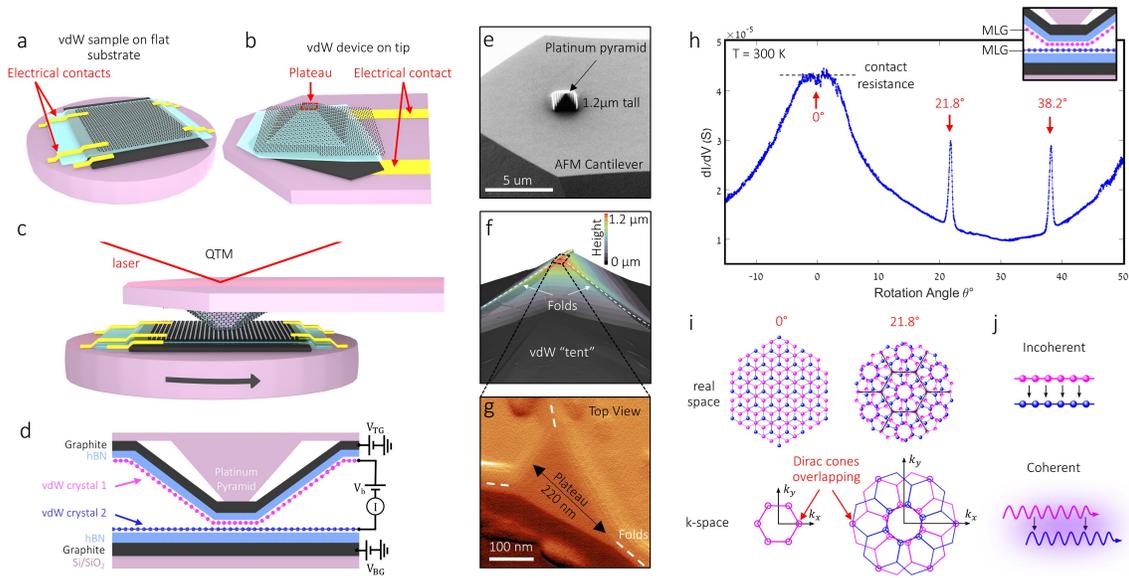

**Fig 1:** T**he Quantum Twisting Microscope setup and in-situ twistronics experiments**. **a**, **b** and **c**, Illustrations of the components of the QTM. **a**. van-der-Waals (vdW) heterostructure assembled on a flat substrate with electrical contacts. **b.** vdW heterostructure assembled on a pyramid positioned near the edge of an atomic force microscope (AFM) cantilever, with electrical contacts. At the apex of the pyramid, the vdW-device-on-tip has a flat plateau. **c.** These devices are brought into contact using a commercial AFM, fitted with a piezoelectric rotator that allows to control the relative angle between tip and sample, θ, continuously with a 0.001° resolution, in addition to the usual lateral (X/Y) scanning capabilities of the AFM. **d**. Measurement circuit plotted over a schematic cross-section of the QTM junction. A voltage bias, $V_b$, is applied between the two active vdW layers (vdW crystals 1 and 2), and the corresponding current, $I$, is measured. Buried top and bottom graphite gate voltages, $V_{BG}$ and $V_{TG}$, can modify the local carrier density and electric field in the junction (in this paper, the top gate is physically shorted to the top graphene layer such that $V_{TG}$ is identically zero**) e**. SEM image of an AFM cantilever with a custom-made platinum pyramid deposited by a focused-ion-beam. **f.** The topography of a vdW-device-on-tip (graphite/hBN/monolayer graphene (MLG)), imaged by AFM. The vdW layers form a tent over the Pt pyramid, with folds (dashed white) leading to a flat plateau. **g.** Zoomed-in AFM image around the pyramid's apex (peak-force-error signal), showing the spontaneously formed flat plateau. **h.** Measured conductance, dI/dV, vs. rotation angle, θ, between two graphene monolayers (MLG, top inset) in direct contact, $V_b = 50mV, V_{BG} = 0, T = 300K$. The two vdW devices are kept in continuous contact throughout the measurement. **i**, Real-space and momentum-space registry for the commensurate angles of 0° and 21.8°. **j**, Illustrations of the incoherent and coherent tunneling across the 2D junction: in the former, electrons tunneling at various locations are incoherent, and the tunneling is proportional only to the local wavefunction squared. In the latter, tunneling trajectories within the coherence length interfere, and the tunneling is sensitive also to the variation of the phases of the top and bottom wavefunctions along the junction.

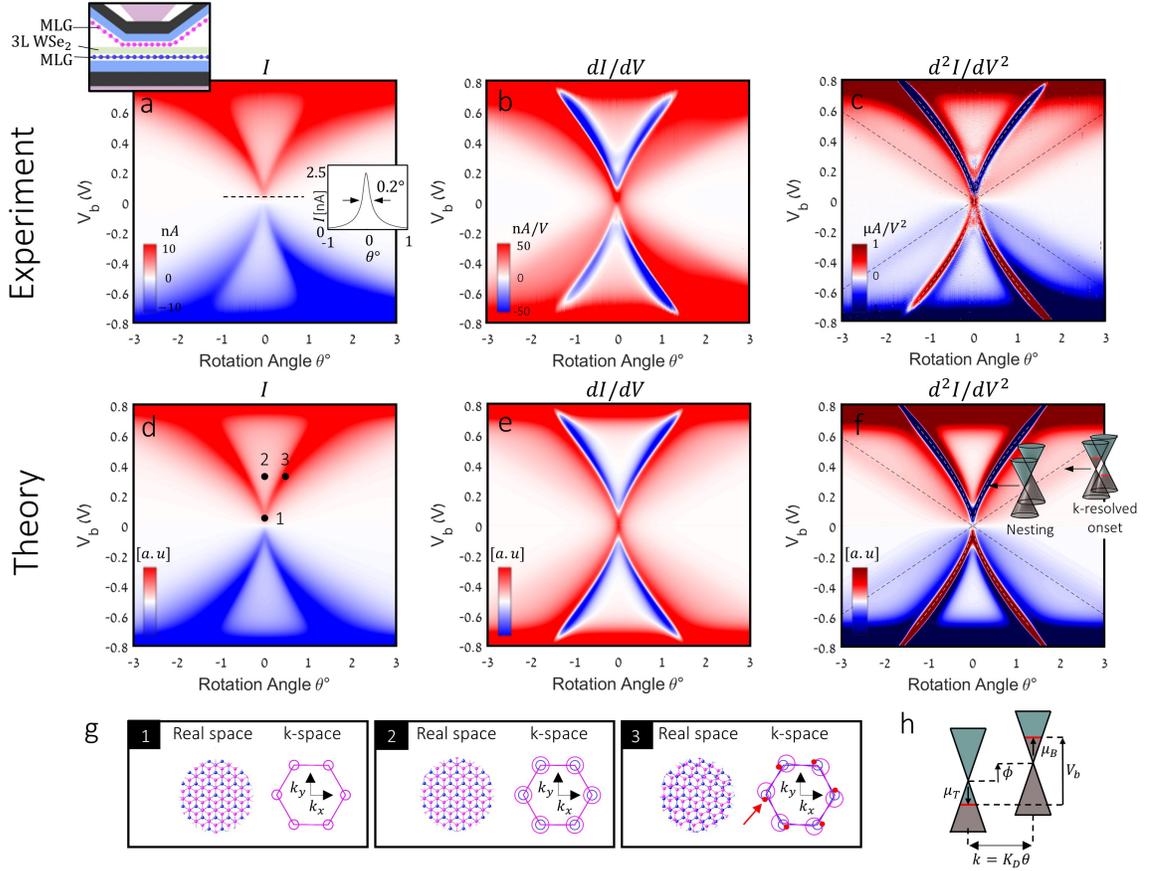

**Fig 2: Momentum resolved tunneling between two twisted graphene monolayers**. **a.** Top inset: Schematic cross-section of the experiment, comprising of an MLG/trilayer $WSe_2$/MLG tunnel junction. The $WSe_2$ tunnel barrier suppresses the hybridization between the MLG layers allowing one to probe the unperturbed energy bands of the other. Main panel: Tunneling current, $I$, measured as a function of rotation angle, $\theta$, and interlayer bias, $V_b$, at $T = 300K$. Right inset: a plot of $I$ vs. $\theta$ at $V_b = 40mV$ (along the black dashed line in the main panel). The peak's full-width-half-max (FWHM) is 0.2°. **b.** A lock-in measurement of the differential conductance, $dI/dV$, as a function of $\theta$ and $V_b$, showing strong negative differential resistance (NDR, blue) along a curved-X feature. The measurement shows a good symmetry between the positive and negative bias directions. **c.** The second derivative, $d^2I/dV^2$, vs. $\theta$ and $V_b$, obtained by a numerical derivative of the data in panel b. The black and white dashed lines correspond to two specific alignment conditions between the bands, 'k-resolved onset' and 'nesting', shown in panel f and described in the main text and Supp. Info. S5 and Supp. Video 4. **d-f.** Theoretically calculated momentum conserving tunneling $I$, $dI/dV$, and $d^2I/dV^2$ between two MLG layers spaced by a tunneling barrier, as a function of $\theta$ and $V_b$, based on the Bistrizer-MacDonald expression for momentum conserving tunneling between rotated vdW layers[23]. The calculation includes finite temperature, $k_BT = 25meV$, and a finite inverse electron lifetime[12], $\gamma = \gamma_0 + \gamma_1|\epsilon - \mu|$, where $\epsilon$ is the electron energy and $\mu$ is the chemical potential. $\gamma_0 = 4meV$, $\gamma_1 = 0.035$ are the experimentally fitted values. The lifetime at low energies corresponds to a coherence length of $l_\phi \approx 150nm$, comparable to the size of our tip, setting a lower bound on the decoherence by electron-phonon and electron-electron interactions at room temperature. The illustrations show the relative alignment of the two Dirac cones at the 'k-resolved onset' and 'nesting' conditions, along the corresponding dashed black and dashed white lines. **g.** panels 1-3 illustrate the relative alignment of the two MLG layers in real space and momentum space, corresponding to points 1-3 in panel d. **h.** Schematic band alignment of the two layers under finite $V_b$ and $\theta$. $V_b$ divides between shifts to the chemical potentials of bottom and top layers, $\mu_B$ and $\mu_T$, and the energy band shift (=electrostatic potential shift), $\phi$. The Dirac points are shifted in momentum by $k = K_D\theta$, where $K_D$ is the momentum of the Dirac point.

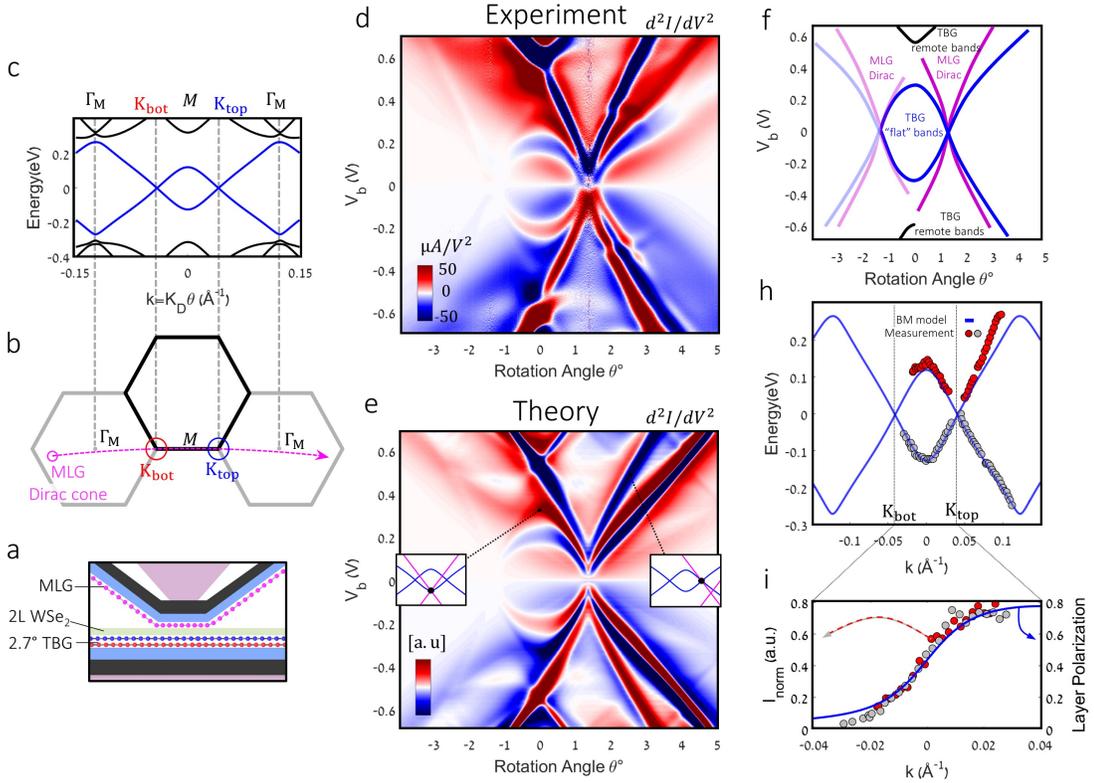

**Fig 3. QTM imaging of the energy bands of twisted bilayer graphene (TBG) a.** Schematics of the experiment, comprised of a MLG/ bilayer WSe$_2$/ 2.7° TBG tunnel junction **b.** The mini-Brillouin zone (BZ) of the TBG, with the Dirac cones of the underlying top and bottom layers (represented by blue and red circles) located at two K-points (K$_{bot}$ and K$_{top}$). Also shown are the $\Gamma_M$ points of adjacent mini-BZs. When the MLG probe layer is rotated with respect to the TBG, its Dirac cone (purple circle) traces the TBG energy bands along an arc in momentum space (dashed purple), crossing through the K$_{bot}$ and K$_{top}$ points, and very close to the $\Gamma_M$ points. **c.** Theoretical energy bands of 2.7° TBG along the dashed purple arc in panel b, calculated using the Bistrizer-MacDonald model[3]. The "flat" and remote bands are shown in blue and black, respectively. **d.** The second derivative, $d^2I/dV^2$, measured as a function of θ and V$_b$ at $T = 300K$. **e.** Theoretically-calculated $d^2I/dV^2$ for this junction, as a function of $\theta$ and $V_b$. The theory is based on the Bistrizer-Macdonald expression for momentum conserving tunneling[3,23] and includes finite temperature, $k_BT = 25 meV$, and finite electron lifetime[12], $\gamma = \gamma_0 + \gamma_1|\epsilon - \mu|$ with $\gamma_0 = 4meV, \gamma_1 = 0.02$. The insets show the relative alignment of the MLG and TBG energy bands at two points in the $\theta$-$V_b$ plane. Left inset: the MLG Dirac point matches in energy and momentum a point on the flat TBG bands. Right inset: the TBG Dirac point matches in energy and momentum a point on the MLG Dirac band. **f,** Tracing of the main features of the measurement in panel d, including features related to the TBG flat (blue) and remote (black) bands, as well as two copies of the MLG Dirac bands (purple). **h.** The TBG energy vs. momentum dispersion (red and grey dots) determined from the measurement in panel d after converting $V_b$ to the energy axis (see text), together with the bands of the BM model (blue). **i**. The magnitude of the tunneling current, $I$, traced along the features that corresponds to the flat bands (blue curves, panel f) between K$_{bot}$ and K$_{top}$ (gray/red points for the positive/negative bias side). We normalize the current values by a constant to compare with the theory. Blue line: the theoretically calculated layer polarization of the wavefunction vs. momentum.

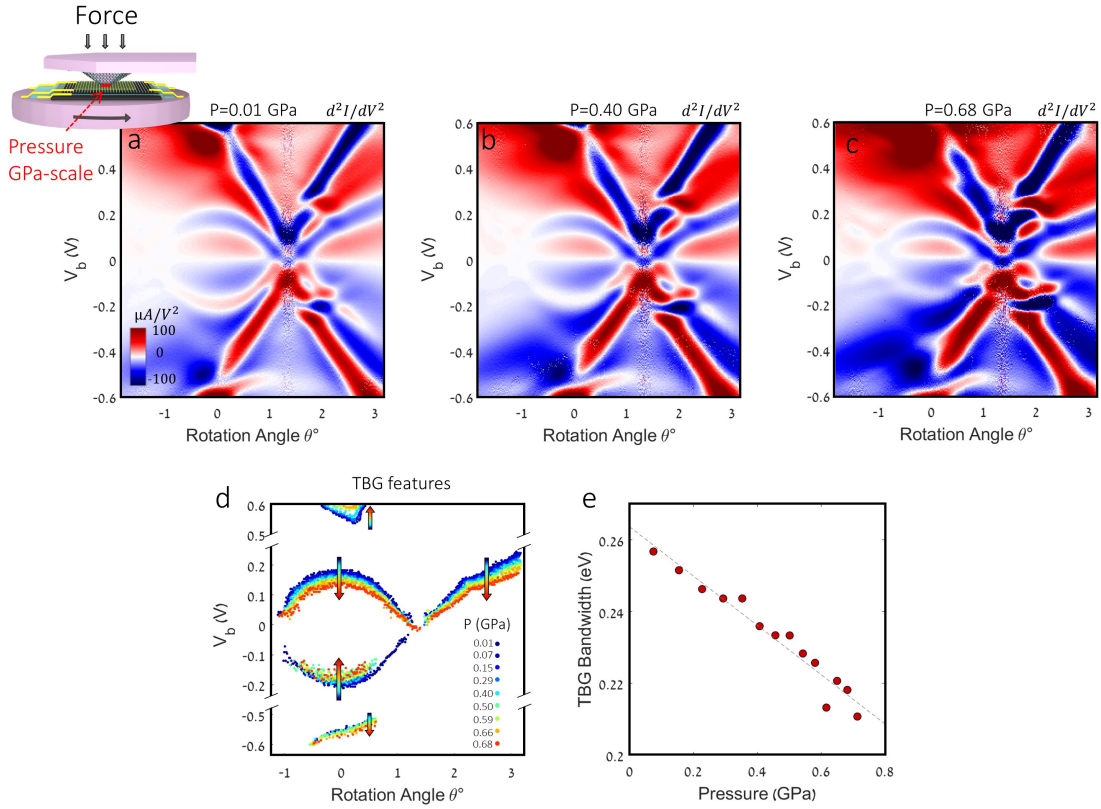

**Fig 4. Imaging the effects of applied pressure on the TBG energy bands: a.** inset: Illustration of the experiment – applying a fixed AFM force across the junction results in GPa-scale pressures at the interface. In the experiment we keep the pressure on while twisting. **a-c.** Measured d²I/dV² vs. θ and $V_b$ for the junction in figure 3, but under applied pressures of P=0.01, P=0.4 and P=0.68 GPa. $T = 300K$. **d**. Evolution of the TBG energy bands with P. In this figure we trace features that reflect the flat and remote bands as a function of pressure (see key). For every pressure, we plot points that were obtained from tracing the relevant feature ('onset' feature for the flat bands, zero $dI^2/dV^2$ contour for the remote bands). Visibly, with increasing $P$, the flat bands gradually shrink toward zero $V_b$ while the remote bands shift to higher $V_b$ (arrows), consistent with a pressure controlled anti-crossing. **e**. The bandwidth at the $M$ point of the TBG "flat" energy bands vs. $P$ (red dots), determined from linecuts at $\theta = 0°$. We convert $V_b$ to the energy from the simultaneously measured MLG Dirac bands (curved-X on the right side in panels a-c, see SI S8). Dashed line is a linear fit to the data. At the highest applied pressure (0.68 GPa), the bandwidth shrinks by ~17%.

**Acknowledgements:** We thank P Jarillo-Herrero, A. Kanigel, Y. Ronen, H. Steinberg and U. Zondiner for useful discussions and comments to the manuscript. Work was supported by the Leona M. and Harry B. Helmsley Charitable Trust grant, the Rosa and Emilio Segre Research Award, the ERC-Cog grant (See-1D-Qmatter, no. 647413) and the BSF grant (2020260).

**Author Contributions:** A.I., J.B, J.X and S.I. designed the experiment. A.I., J.B and J.X built the setup, fabricated the devices, and performed the experiments. A.I., J.B, J.X and S.I. analyzed the data. J.X, B.Y.,Y.O., A.S. and E.B. wrote the theoretical models K.W. and T.T. supplied the hBN crystals. A.I., J.B, J.X, A.S. and S.I. wrote the manuscript with input from other authors.

**Data availability:** The data shown in this paper is provided with this paper. Additional data that support the plots and other analysis in this work are available from the corresponding author upon request.

**Competing interests:** The authors declare no competing interests